\documentstyle[12pt]{article}
\newcommand{\varPhi}{\Phi}
\newcommand{\drm}{{\rm d}}
\newcommand{\bi}{\bibitem}
\newcommand{\til}{\tilde}

\newcommand{\Lc}{{\cal L}}
\newcommand{\Hc}{{\cal H}}

\newcommand{\Fc}{{\cal F}}

\newcommand{\om}{\omega}
\newcommand{\be}{\begin{equation}}
\newcommand{\ee}{\end{equation}}
\newcommand{\bea}{\begin{eqnarray}}
\newcommand{\eea}{\end{eqnarray}}
\newcommand{\beas}{\begin{eqnarray*}}
\newcommand{\eeas}{\end{eqnarray*}}

\newcommand{\ga}{\gamma}

\newcommand{\erm}{{\rm e}}
\newcommand{\gabf}{\mbox{\boldmath $\gamma$}}

\newcommand{\dar}{\dagger}

\newcommand{\sig}{\sigma}

\newcommand{\vphi}{\varphi}
\newcommand{\ria}{\rightarrow}
\newcommand{\longra}{\longrightarrow}
\newcommand{\upa}{\uparrow}

\newcommand{\al}{\alpha}

\newcommand{\veps}{\varepsilon}

\newcommand{\1}{1\!\!1}
\newcommand{\lan}{\langle}

\newcommand{\ran}{\rangle}
\newcommand{\Om}{\Omega}
\newcommand{\th}{\theta}

\newcommand{\R}{I\!\!R}

\newcommand{\h}{\hspace*{0.5 cm}}
\newcommand{\dis}{\displaystyle}

\newcommand{\psit}{\tilde{\psi}}
\newcommand{\bt}{\beta}

\newcommand{\cent}{\centerline}
\newcommand{\vs}{\vspace*}

\newcommand{\r}{\rho}
\newcommand{\dpar}{\partial}
\newcommand{\pa}{\dpar}
\newcommand{\zb}{\bar{z}}

\newcommand{\doz}{\dot{z}}

\newcommand{\dozb}{\dot{\bar{z}}}

\newcommand{\dox}{\dot{x}}
\newcommand{\dopi}{\dot{\pi}}
\newcommand{\dopsi}{\dot{\psi}}

\begin{document}
 
\cent{{\bf ELECTRON STRUCTURE, ZITTERBEWEGUNG AND A NEW}}
\cent{{\bf NON-LINEAR DIRAC--LIKE EQUATION$^{\: \dag}$}}

\vs{1.5cm}

\cent{{Matej PAV\v{S}I\v{C}$^{*}$, \ Erasmo RECAMI$^{**}$, \  Waldyr A. 
RODRIGUES Jr.$^{***}$}}

\vs{0.5 cm}
  
\cent{{\em I.N.F.N. -- Sezione di Milano, Milan, Italy.}}

\footnotetext{$^{\: \dag}$ Work partially supported by INFN, CNR, MURST;
by CAPES, CNPq, FAPESP; and by the Slovenian Ministry of Science and 
Technology.} 
\footnotetext{$^{*}$ On leave from the J.Stefan Institute; University of
Ljubljana; 61111--Ljubljana; Slovenia.}
\footnotetext{$^{**}$ Also: Facolt\`a di Ingegneria, Universit\`a statale 
di Bergamo, 24044 Dalmine (BG), Italy; \ and C.C.S.,
State University at Campinas, 13083-970 Campinas, S.P.; Brazil.}
\footnotetext{$^{***}$ On leave from Departamento de Matem\'{a}tica Aplicada 
 --- Imecc; UNICAMP; 13084--Campinas, S.P.; Brazil.}

\vs{2cm}
 
{\small\bf ABSTRACT:}  The recent literature shows a renewed interest, with
various independent approaches, in the classical theories for spin. 
Considering the possible interest of those results, at least for the
electron case, we purpose in this
paper to explore their physical and mathematical meaning, by the natural and
powerful language of Clifford algebras (which, incidentally, will allow us to
unify those different approaches). \  In such theories,
the ordinary electron is in general
associated to the mean motion of a point--like ``constituent" $\cal Q$, whose
trajectory is a cylindrical helix. \ We find, in particular, 
that the object $\cal Q$ obeys a new, non-linear
Dirac--like equation, such that ---when averaging over an internal cycle
(which corresponds to
linearization)--- it transforms into the
ordinary Dirac equation (valid for the electron as a whole).
 
\vspace*{2.0 cm}


\vfill
 
\newpage


{\bf 1. \ INTRODUCTION}\\ 

\h The possibility of constructing formal
classical theories for spin was already realized at least 60 years ago, from 
different points of view.$^{1}$ \ In particular, Schr\"odinger's 
suggestion$^{2}$ that the electron spin
was related to its Zitterbewegung (zbw) motion has
been investigated by several authors$^{3}$. 

\h In ref.$^{4}$, for instance,
one meets even the proposal of models with 
clockwise and anticlockwise ``inner motions"  as classical analogues 
of quantum relativistic spinning particles
and antiparticles, respectively. \ The use of grassmannian variables in a 
classical 
lagrangian formulation for spinning particles was proposed, on the contrary, 
by Berezin and Marinov$^{5}$ and
by Casalbuoni$^{6}$. \ Pauri,$^{7}$ moreover,  showed how a
relativistic  (or non-relativistic) spin can enter
a classical formulation, in the ordinary phase space [without
resource now to grassmannian quantities], just as a consequence of the
algebraic structure of the Poincar\'e (or Galilei) group. In such an
interesting approach, it was found$^{7}$ that a relativistic classical
zbw motion can directly follow as a ``spin effect" from the requirements
of existence of a lagrangian and of a covariant position in Dirac's
instant form of relativistic dynamics. The quantum analogue of those
developments had been studied in ref.$^{8}$. 

\h The number  of papers
appeared on the subject of classical theories for spin, starting from the
fifties, is so large that it would be difficult to try quoting them here.
 \ A recent approach,
based on a generalization of Dirac non-linear electrodynamics,
where the spin of the electron is identified with the momentum of the
Poynting vector in a soliton solution of that theory, can be found
in ref.$^{9}$.
 
\h In this paper we choose, by making recourse to the natural and powerful
language of Clifford algebras,  to refer ourselves mainly to 
the theory by Barut and Zanghi$^{10,11}$ (BZ), \  which relates the spin 
(at least in
the case of the electron) to a helical motion. \ Namely, we first recast
the BZ theory into the Clifford formalism, in the meantime clarifying
its physical and mathematical meanings; \ then, we quantize that theory for
the electron case. \ In particular, we derive from the BZ theory a 
{\em new} non-linear
equation for the ``spinorial variable" of the model, which  when
linearized reduces to the ordinary Dirac
equation. \ Solutions of this non-linear equation will be discussed
in another paper$^{12}$.
 
\vspace*{1.5 cm}

{\bf 2. \ SPIN AND ELICAL MOTION}\\  

\h In the BZ theory,$^{10}$ 
a classical electron is
characterized, besides by the usual pair of conjugate variables
($x^{\mu},p_{\mu}$), by a second pair of conjugate classical 
{\em spinor} variables ($z,i\bar{z}$), representing internal degrees of
freedom, which are functions of an invariant time parameter $\tau$, that 
when convenient will be identified with the proper time
of the center of mass (CM).  Quantity $z$ was a Dirac spinor, while 
${\bar z} \equiv z^{\dar} \gabf^{0}$. \ Barut and Zanghi, then, introduced 
for a spinning particle the classical lagrangian [$c = 1$]

\

\hfill{
${\cal{L}}  \; =  \; {1 \over 2} \lambda i ( {\dot{\bar{z}}} z - \bar{z} 
\dot{z} )  \;
+  \; p_{\mu} ({\dot{x}}^{\mu} - \bar{z}   \gabf^{\mu} z)  \; +  \;
e A_{\mu}(x) \bar{z} \gabf^{\mu} z \; \; \; ,$                      
\hfill} (1)                                                   

\

where $\lambda$ has the dimension of an action, $\gabf^{\mu}$ are the Dirac
matrices, and the particle velocity is

\

\hfill{
$v_{\mu} \equiv  \bar{z} \gabf_{\mu} z \; \; .$ 
\hfill} (1')

\

We are not  writing  down explicitly the spinorial indices of $\bar{z}$ and 
$z$. \ Let us consider the simple case of a {\em free} electron 
($A_{\mu} = 0$). Then a possible solution of the equations of motion 
corresponding to the lagrangian (1) is:$^{\#1}$
\footnotetext{$^{\#1}$ For other solutions, see ref.$^{12}$.}

\

\hfill{$  z(\tau) \; \;  =  \; \; [\cos  m\tau \; - \; i\gabf^{\mu} 
\dis{{p_{\mu} \over m}} \sin m\tau] \; z(0) \; \; ,$ \hfill} (2a)

\

\hfill{$  \bar{z}(\tau) \; \;  =  \; \; \bar{z}(0) \; [\cos m\tau \; + \; 
i\gabf^{\mu} \dis{{p_{\mu}
\over m}} \sin m\tau] \; \; ,$ \hfill} (2b)           
      
\
                                                   
and \ $p_{\mu} = {\rm constant}$; \ $p^{2} = m^{2}$; \ $H = p_{\mu} \bar{z} 
\gabf^{\mu} z  \equiv p_{\mu}v^{\mu}$; \ and finally:

\

\hfill{$ {\dot{x}}_{\mu}  \; =  \; v_{\mu}  \; =  \; \dis{{p_{\mu} \over
m^{2}}} H  \; +  \; [{\dot{x}}_{\mu}(0) \; - \; \dis{{p_{\mu} \over
m^{2}}} H] \cos 2m\tau \;  +  \; \dis{{{\ddot{x}}_{\mu}(0) \over 2m}} 
\sin 2m\tau   $ \hfill} (2c)

\

(which in ref.$^{10}$ appeared with two misprints). In connection with 
this ``free" general solution, let us remark that $H$
is a constant of motion so that we can set $H = m$. \ Solution (2) exhibits the 
classical analog of the phenomenon known as ``zitterbewegung" (zbw): \ in 
fact, the velocity $v_{\mu}
\equiv \dot{x}_{\mu}$ contains the (expected) term $p_{\mu}/m$ plus a 
term describing an oscillatory motion with the characteristic frequency 
$\omega = 2m$. \ The velocity of the {\em center of mass} will be given by 
$W_{\mu} = 
p_{\mu}/m$. \ Notice incidentally that, instead of adopting the variables $z$
and $\bar{z}$, one can work in terms of the spin variables, {\em i.e.} in
terms of the dynamical variables $(x_{\mu}, v_{\mu}, \pi_{\mu}, S_{\mu \nu})$, 
where:

$$v_{\mu} = \dot{x}_{\mu}; \;\; \pi_{\mu} = p_{\mu} - eA_{\mu}; \; \; \; \;
S_{\mu \nu} = {1 \over 4} i \bar{z} [\gabf_{\mu}, \gabf_{\nu}] z \; ,$$

\

so that ${\dot{S}}_{\mu \nu} = \pi_{\mu} v_{\nu} - \pi_{\nu} v_{\mu}$ \ and \ 
${\dot{v}}_{\mu} = 4 S_{\mu \nu} \pi^{\nu}$. \ In the case of a free electron, 
by varying the action corresponding to ${\cal{L}}$ one finds as generator of 
space-time rotations the conserved quantity $J_{\mu \nu} = L_{\mu \nu} + 
S_{\mu \nu} \;$, \ where \ $S_{\mu \nu}$ is just the particle spin.$^{\#2}$
\footnotetext{$^{\#2}$ Alternatively, Barut and Zanghi,$^{10}$ in order to study
the internal dynamics of the considered (classical) particle, did split 
$x_{\mu}$ and $v_{\mu} \equiv {\dot x}_{\mu}$ as follows: \ $x_{\mu} \equiv
X_{\mu} + Q_{\mu}; \; v_{\mu} \equiv W_{\mu} + U_{\mu}$ (where by
definition $W_{\mu} = {\dot X}_{\mu}$ and $U_{\mu} = {\dot Q}_{\mu}$). \ In 
the particular case of a 
{\em free} particle, ${\dot W}_{\mu} = 0; \; W_{\mu} = p_{\mu}/m$. \ One can
now interpret $X_{\mu}$ and $p_{\mu}$ as the CM coordinates, and $Q_{\mu}$
and $P_{\mu} \equiv mU_{\mu}$ as the {\em relative} position and momentum, 
respectively. For a free particle, then, one finds that the internal variables
are coordinates oscillating with the zbw frequency $2m$; and that, again, the
total angular momentum $J_{\mu \nu} = L_{\mu \nu} + S_{\mu \nu}$ is a constant
of motion, quantities $S_{\mu \nu}$ being the spin variables.}

\h Let us explicitly observe that solution (2c) is the equation of a
space-time cylindrical helix, {\em i.e.} it represents in 3--space a helical
motion. Let us also stress that this  motion describes (when one makes recourse to the
light-like solutions: see the following)
particle spin at a classical level. In fact, such a classical system has 
been shown to 
describe, after quantization, the Dirac electron. \ Namely, Barut and 
Pav\v{s}i\v{c}$^{13}$ started from the classical hamiltonian corresponding to 
eq.(1):

\

\hfill{$\Hc \; = \; {\bar z} \gabf^{\mu} z (p_{\mu} - eA_{\mu}) \; ;$ 
\hfill} (3)                                                      

\

passed to its quantum version, in which the above quantities are regarded
as operators; and considered in the Schroedinger picture the equation

\

\hfill{$ i \dis{{{\dpar \varphi} \over {\dpar \tau}}} \; = \; \Hc \varphi $ 
\hfill} (4)  

\ 

where $\vphi = \vphi (\tau, x, z)$ is the wave function, and the operators in
eq.(3) are \ $p_{\mu} \longra -i {{\dpar } / {\dpar x^{\mu}}}$ \ and \ 
$i {\bar z} \longra  -i {{\dpar } / {\dpar z}}$. \ The wave function $\vphi$
can be expanded in the $z$ variable, around $z = 0$, as follows:

\

\hfill{$\vphi (\tau, x, z) \; = \; {\bar \varPhi}(\tau, x) \; + \; 
{\bar \psi}_{\al}(\tau, x) z^{\al} \; +  \; {\bar \psi}_{\al \beta}(\tau, x)
z^{\al} z^{\beta} \; + \;$...  \hfill}

\

where  $\al$, $\bt$ are spinorial indices. Quantity \  
${\bar \psi} \equiv {\psi^{\dar}} \gabf^{0}$ \ is the Dirac adjoint spinor,
which a charge ${\bar e}$ and a mass ${\bar m}$ are attributed to. \ By
disregarding the spin--zero term ${\bar \varPhi}(\tau, x)$, \ and retaining
only the second (spin ${1 \over 2}$) term, thus   neglecting also the
higher--spin  terms, from eq.(4) they then derived the equation 

\

\hfill{$i \dis{{{\dpar {\bar \psi}} \over {\dpar \tau}}} \; = \; 
(i{\dpar }_{\mu}
+ {\bar e} A_{\mu}) {\bar \psi} \gabf_{\mu} \; \; .$  \hfill} (5)
                                        
\

Taking the Dirac adjoint of such an equation, with$^{14}$ \ ${\bar e} = -e$;
${\bar m} = -m$, \ one ends up just with the Dirac equation. This confirms,
by the way, that the term ${\bar \psi}_{\al} z^{\al}$ refers to the
spin ${1 \over 2}$ case, i.e. to the case of the electron.$^{\#3}$
\footnotetext{$^{\#3}$ For the case of higher spins, cf. ref.$^{15}$.} 

\

\h An alternative approach leading to a classical description of particles 
with spin is the one by Pav\v{s}i\v{c},$^{16,17}$  who made recourse to the 
(extrinsic)
curvature of the particle world--line in Minkowski space; so that his 
starting, classical lagrangian [$\al$,$\bt$ = constants]

\

\hfill{${\Lc} \; = \; \sqrt{{\dot x}^{2}} \; (\al - \bt K^{2}); \; \; K^{\mu}
\; \equiv \; \dis{{1 \over \sqrt{{\dot x}^{2}}} \; {{\rm d} \over {\rm d}\tau} 
\left({{\dot x}^{\mu} \over \sqrt{{\dot x}^{2}}} \right)}$
\hfill} (6)

\

contained the extra (kinematical) term  \ $\bt {\sqrt{{\dot x}^{2}}} \: K^{2}$.
 \ The conserved generator of rotations, belonging to lagrangian (6), is once
more  \ $J_{\mu \nu} = L_{\mu \nu} + S_{\mu \nu} \;$, \  
where now, however, \ $S_{\mu \nu} = {\dot x}_{\mu} p^{(2)}_{\nu} - 
{\dot x}_{\nu} p^{(2)}_{\mu}$, \ while \ $p^{(2)}_{\nu} \: \equiv \: 
{{\dpar \Lc} / {\dpar {\ddot x}^{\mu}}} \: = \: 2\bt {\ddot x}_{\mu} /
{\sqrt{{\dot x}^{2}}}$ \ is the second--order canonical momentum, conjugated 
to ${\dot x}^{\mu}$. \ The equations of motion, which correspond to eq.(6) 
for constant
$K^{2}$, do again admit as solution$^{16,17}$ the {\em helical} motion with 
the zbw frequency $\om = 2m$. \  For a suitable choice of the constant $K^{2}$,
and when the affine parameter $\tau$ is the CM proper--time, the equations
of motion result to be [$v_{\mu} \equiv {\dot x}_{\mu}$]

\

\hfill{${\dot v}_{\mu} \: = \: 4 S_{\mu \nu} p^{\nu}; \; \; 
{\dot{S}}_{\mu \nu} \: = \: \pi_{\mu} v_{\nu} - \pi_{\nu} v_{\mu} \; ,$
\hfill}

\

{\em i.e.}, they are the same (for the free case: $A_{\mu} = 0$) as in the 
Barut--Zanghi (BZ)
theory. \ Moreover, the constraint  due to reparametrization invariance can
be written as \ $p_{\mu} v^{\mu} - m~=~0$, which reminds us of the Dirac
equation; and Poisson brackets are obtained which obey the same 
algebra as
the Dirac $\gabf$ matrices.$^{\#4}$ \ Actually, the velocity
$v_{\mu}$ can be also expressed, in terms of the canonically conjugate
variables ${\dot{x}}^{\mu}$ and $p^{(2)}_{\mu}$, as follows:

\

\hfill{
$v^{\mu} \; = \; k {\dis{{{\dot x}^{\mu} \over {\sqrt {{\dot x}^2}}}}} \; ; 
\;\;\;\;\;\;\;\;\; k \equiv {\dis{{4 \bt m} \over {4 \alpha \bt + 
{{p^{(2)}_{\mu} {p^{(2)}}^{\mu}}}}}} \; .$      
\hfill}

\

Then, in ref.$^{16}$ there were obtained the Poisson brackets:

\

\hfill{$\lbrace v^{\mu}, v^{\nu} \rbrace = 4 S^{\mu \nu} \; ;$ 
\hfill} 

\

\hfill{$\lbrace S^{\mu \nu}, v^{\rho} \rbrace = g^{\mu \rho} v^{\nu} - 
g^{\nu \rho} v^{\mu} \; ,$
\hfill} 

\

where $k$ was chosen in such a way that $k^3 / (2 \bt m) = 4 \,$.

\h The above relations 
are analogous to the corresponding quantum equations, in which $v^{\mu}$
is replaced by the Dirac operators ${\gabf}^{\mu}$, and the Poisson brackets
by commutators. The classical frequency of the orbital motion is equal
to the zbw frequency of the Dirac theory.
\footnotetext{$^{\#4}$ This conclusion, referring to a half--integer spin, 
has been criticized in refs.$^{18}$ by noticing a
seeming analogy with the case of the ordinary orbital angular
momentum (which admits integer spin values only). We believe such arguments
to not apply to the present situation, where spin is the orbital momentum
in the $q^{\mu} \equiv {\dot x}^{\mu}$ variables space; since the 
corresponding, canonically conjugate momentum $p^{(2)}_{\mu}$ is never a 
constant
of motion. [Namely, when we consider a congruence of world-lines which are
solutions of the equations of motion, one finds a non--zero curl of the field 
$p^{(2)}_{\mu}(q)$, so that the phase \ $\int p^{(2)}_{\mu} {\rm d}q^{\mu}$ \ 
is not a single--valued function. Thus, a basis of single--valued wave 
functions does not exist, nor the operator representation \ $p^{(2)}_{\mu}
\longra + i {\dpar } / {\dpar q^{\mu}}$; \ and the ordinary arguments about
orbital angular momentum do no longer apply].}

\

\h Let us finally mention that the {\em same} classical equations of motion 
(and the same Poisson--bracket algebra) have been found also in a third 
approach, which consists in adding to the ordinary lagrangian an extra term
containing Grassmann variables.$^{19}$  

\vspace*{1.5 cm}

{\bf 3. \ ABOUT THE ELECTRON STRUCTURE}\\   

\h Considering the interest 
of the previous results
(which suggest in particular that the helical motion can have a role in the
origin  of spin),
we purpose to explore their physical meaning more deeply, by the very natural
---and powerful--- language of the Clifford algebras:$^{20,21}$ \  in 
particular of the ``space-time algebra (STA)" ${\R}_{1,3}$. \ First of all, let
us preliminarily clarify {\em why} Barut and Zanghi had to introduce the Dirac
spinors $z$ in their lagrangian, by recalling that classically the motion of a
spinning top has to be individuated by \ (i) the world--line $\sig$ of its 
center of mass ({\em e.g.}, by the ordinary coordinates $x^{\mu}$ and the 
conjugate momenta $p_{\mu}$), and \ (ii) a Frenet tetrad$^{\#5}$
\footnotetext{$^{\#5}$ The use of Frenet tetrads in connection with the Dirac 
formalism was first investigated in ref.$^{22}$.}
attached$^{22}$ to the world--line $\sig$. \ This continues to be true when
wishing to describe the motion of a point-like spinning particle. \ For the 
Frenet tetrad$^{23}$ we have:
\

\hfill{$e_{\mu} \: = \: R \ga_{\mu} {\til{R}} \: = \: \Lambda_{\mu}^{\nu} 
\ga_{\nu} \: ; \; \; \; \Lambda_{\mu}^{\nu} \in L_{+}^{\upa}$ \hfill} (7)

\

where $e_{0}$ is parallel to the particle velocity $v$ (even more, $e_0 = v$ 
whenever one does use as parameter $\tau$ the CM system proper--time); 
the tilde represents the {\em reversion}$^{\#6}$;  
\footnotetext{$^{\#6}$ \h The main anti-automorphism in $\R_{1,3}$ 
(called reversion), denoted by the tilde, is such that   
$\widetilde{AB} = \til{B} \til{A}$, and $\til{A} = A$ when $A$ is a
scalar or a vector, while $\til{F}=-F$ when $F$ is a 2-vector.}
and $R = R(\tau)$ is a ``Lorentz rotation" [more precisely, \ $R \: \in \: 
{\rm Spin}_{+}(1,3)$, and a Lorentz transform of quantity $a$ is 
given by $a' = R a \til{R}$]. \ Moreover \ $R \til{R} \: = \: 
\til{R} R \: = \: 1 \:$. \ The Clifford STA fundamental
unit--vectors $\ga_{\mu}$  should not be confused with the Dirac 
{\em matrices} $\gabf_{\mu}$.
Let us also recall that, while the orthonormal vectors $\ga_{\mu} \equiv 
{\pa / {\pa x^{\mu}}}$ constitute a {\em global} tetrad in Minkowski 
space-time 
(associated with a given inertial observer), on the contrary the Frenet
tetrad $e_{\mu}$ is defined only along $\sig$, in such a way that $e_0$
is tangent to $\sig$. \ At last, it is: \ $\ga^{\mu} = 
\eta^{\mu \nu} \ga_{\nu}$, \ and \ $\gamma_5 \equiv \ga_0 \ga_1 \ga_2
\ga_3$.\\
\h Notice that $R(\tau)$ does contain all the essential information carried
by a Dirac spinor. In fact, out of $R$, a ``Dirac--Hestenes" (DH) 
spinor$^{24}$ \  $\psi_{\rm DH}$ \ can be constructed as follows:

\

\hfill{$\psi_{\rm DH} \; = \; \r^{1 \over 2} \erm^{{\bt \gamma_{5}} /2} R$ 
\hfill} (8)

\

where $\r$ is a normalization factor; and \ $\erm^{{\bt \gamma_{5}}} = 
+1$ \ for the electron (and $-1$ for the positron); while, if $\veps$ is a 
primitive idempotent of the STA, any Dirac spinor $\psi_{\rm D}$ can be 
represented in our STA as:$^{25}$
 
\
                                                                               
\hfill{$\psi_{\rm D} \; = \; \psi_{\rm DH} \, \veps \: .$ \hfill} (9)

\

For instance, the Dirac spinor $z$ introduced by BZ is obtained from the DH
spinor$^{\#7}$ by the choice \ $\veps \: = \: {1 \over 2} (1 + \ga_{0})$. \ 
\footnotetext{$^{\#7}$ The DH spinors can be regarded as the {\em parent} 
spinors, since all the other spinors of common use among physicists 
are got from them by operating as in eq.(9). We might
call them ``the {\em fundamental} spinors".} 
Incidentally, the Frenet frame can also write \ $\r e_{\mu} \: =
 \: \psi_{\rm DH} \ga_{\mu} {\psit_{\rm DH}}$.\\  
\h Let us stress that, to specify how does the Frenet tetrad rotate as $\tau$
varies, one has to single out a particular $R(\tau)$, and therefore a DH
spinor $\psi_{\rm DH}$ , and eventually a Dirac spinor $\psi_{\rm D}$. \ This
makes intuitively clear why the BZ Dirac--spinor $z$ can a priori provide a 
description of the ``spin motion" of a classical particle.\\ 
\h Let us now repeat what precedes  on a more formal ground. \ In the following,
unless differently stated, we shall indicate the DH spinors $\psi_{\rm DH}$ 
simply by $\psi$.\\

\h  Let us translate the BZ lagrangian into the Clifford language.\\
\h In eq.(1) quantity $z^{\rm T} = (z_1 \;\; z_2 \;\; z_3 \;\; z_4)$ is a 
Dirac spinor $\tau \longra z(\tau)$, and $\bar{z} = z^{\dar} \gabf^{0}$.
To perform our translation, we need a matrix representation of the Clifford 
space-time algebra; this can be implemented by {\em representing} the 
fundamental
Clifford vectors \ ($\ga_0$, $\ga_1$, $\ga_2$, $\ga_3$) \ by the 
{\em ordinary} Dirac matrices $\gabf_{\mu}$.  Choosing:

$$\ga_0 \longra \gabf_0 \: \equiv \: \left( \begin{array}{cc} \1 & 0 \\ 0 & -\1 
\end{array} \right)
\; ; \;\;\;\;  \ga_i \longra \gabf_i \: \equiv \: \left( \begin{array}{cc} 
0 & -\sig_i \\ \sig_i & 0  \end{array} \right) \; ,$$

the representative in  $\R_{1,3}$ of Barut--Zanghi's quantity $z$ is
                                    
\

\hfill{$z \longra \Psi \equiv \psi \, \veps \; ; \;\;\; \veps \equiv {1 \over 2}
(1 + \ga_0)$
\hfill} (10)

\

where $\psi$, and $\psit$, are represented \ (with $\psi \psit = \1$) \ by

\

\hfill{$\psi \; = \; \left( \begin{array}{cccc} z_1 & -\zb_2 & z_3 & \zb_4 \\
z_2 & \zb_1 & z_4 & -\zb_3 \\ z_3 & \zb_4 & z_1 & -\zb_2 \\ z_4 & -\zb_3 & 
z_2 & \zb_1 \end{array} \right) \; ; \;\;\; \psit \; = \; \left( \begin{array}
{cccc} \zb_1 & \zb_2 & -\zb_3 & -\zb_4 \\
-z_2 & z_1 & -z_4 & z_3 \\ -\zb_3 & -\zb_4 & \zb_1 & -\zb_2 \\ -z_4 & z_3 & 
-z_2 & z_1 \end{array} \right) \; .$ 
\hfill} (11)

\
     
The translation of the various terms in eq.(1) is then:

\

\hfill{$\begin{array}{lcl} {1 \over 2}i(\dozb z - z\doz) & \longra & \lan 
\psit \dopsi \ga_1 \ga_2 \ran_0  \\ p_{\mu}(\dox^{\mu} - \zb \gabf^{\mu} z)
&  \longra & \lan p(\dox - \psi \ga_0 \psit ) \ran_0  \\ e A_{\mu} \zb
\gabf^{\mu} z & \longra & e \: \lan A \psi \ga_0 \psit \ran_0 \; , \end{array}$
\hfill}

\

where $\lan \;\;\; \ran_0$ means ``the scalar part" of the Clifford product.
Thus, the lagrangian $\Lc$ in the Clifford formalism is

\

\hfill{$\Lc \; = \; \lan \psit \dopsi \ga_1 \ga_2 \: + \: p(\dox - 
\psi \ga_0 \psit) \: + \: eA \psi \ga_0 \psit \ran_0 \; ,$
\hfill} (12)

\

which is analogous, incidentally, to Kr\"{u}ger's lagrangian$^{26}$ (apart 
from a misprint).\\
\h As we are going to see, by ``quantizing" it, also in the present formalism 
it is possible (and,
actually, quite easy) to derive from $\Lc$ the Dirac--Hestenes 
equation:$^{\#8}$

\

\hfill{$\dpar \, \psi (x) \, \ga_1 \ga_2 \; + \; m \, \psi (x) \, 
\ga_0 \; + \; e \, A(x) \, \psi (x) \; = \; 0 \; ,$
\hfill} (13)

\

\footnotetext{$^{\#8}$ Observe that in eq.(12) it is  $\psi = \psi (\tau)$, 
while in
eq.(13) we have $\psi = \psi (x)$  with $\psi (x)$ such that its 
{\em restriction} \ ${\psi (x)}_{\mid \sig}$ \  
to the world--line $\sig$ coincides with $\psi (\tau)$. \ Below, we shall 
meet the same 
situation, for instance,  when passing from eq.(14a) to eqs.(16)--(16').}
which is nothing but the ordinary Dirac equation written down in the Clifford
formalism.$^{20,24}$ \ Quantity $\dpar \: = \: \ga^{\mu} \dpar_{\mu}$ is the 
Dirac operator. \  Let us notice that $p$ in eq.(12) can be regarded as a 
Lagrange
multiplier, when the velocity $v = \dox$ is represented by $\psi \ga_0 \psit$.
 \ The BZ theory is indeed a hamiltonian system, as proved by using Clifford
algebras in ref.$^{27}$ (cf. also ref.$^{28}$). \ 
The dynamical variables are then ($\psi , \psit , x, p$), and the 
Euler--Lagrange equations yield a system of three independent equations:  

\

\hfill{$\dopsi \ga_1 \ga_2 + \pi \psi \ga_0 \; = \; 0 $
\hfill} (14a)

\hfill{$\dox \; = \; \psi \ga_0 \psit $          
\hfill} (14b)

\hfill{$\dopi \; = \; e F \cdot \dox $
\hfill} (14c)                                                 

\
 
where \ $F \equiv \pa \wedge A$ \ is the electromagnetic field (a bivector, 
in Hestenes' language) \ and \ $\pi \equiv p - eA$ \ is the kinetic 
momentum. \ [Notice incidentally, from eq.(14b), that $\dox^2 = 
\r (\tau)$].\\
 \h At this point, let us consider a velocity vector {\em field} $V(x)$ 
together with 
its integral lines (or stream--lines).  Be $\sig$ the stream--line along which
a particle moves ({\em i.e.}, the particle world--line). Then, the velocity
distribution $V$ is required to be such that its restriction \ 
${V(x)}_{\mid \sig}$ \
to the world--line $\sig$ is the ordinary velocity $v = 
v(\tau)$ of the considered particle.\\ 
\h If we moreover recall$^{29,30}$ that any Lorentz ``rotation" $R$
can be written \ $R = \erm^{\Fc}$, \ where $\Fc$ is a {\em bivector}, 
then along any stream--line $\sig$ we shall have:$^{19}$ 

\

\hfill{$\dot{R} \equiv \dis{{\drm R \over {\drm \tau}}} \; = 
\; {1 \over 2} v^{\mu} \Om_{\mu} R \; = \; {1 \over 2} \Om R \; ,$
\hfill} (15)

\

with \ $\pa_{\mu} R \, = \, \Om_{\mu} R/2$, \ where $\Om_{\mu} \equiv 2 
\pa_{\mu} \Fc$, and where $\Om \equiv v^{\mu} \Om_{\mu}$ 
is the angular--velocity bivector (also known, in differential
geometry, as the ``Darboux bivector"). \ Therefore, 
for the tangent vector {\em along any line} $\sig$ we obtain the relevant 
relation: 

$${\drm \over {\drm \tau}} \; = \; v^{\mu} \pa_{\mu} \; = \; v \cdot \pa \; 
$$

The [total derivative] equation (14a) thus becomes:$^{\#8}$

\

\hfill{$v \cdot \pa \psi \ga_1 \ga_2 + \pi \psi \ga_0 \; = \; 0 \; ,$
\hfill} (16)

\

which is a non-linear [partial derivative] equation, as it is easily seen 
by using eq.(14b) and rewriting it in the noticeable form

\

\hfill{$(\psi \ga_0 \psit) \cdot \pa \psi \ga_1 \ga_2 + \pi \psi \ga_0 \; = 
\; 0 \; .$
\hfill} (16')

\

Equation (16') constitutes a new {\em non-linear} Dirac--like equation.
 \ The solutions of this new
equation will be explicitly discussed elsewhere.\\
\h Let us here observe only that the probability current $J \equiv V$ is
conserved: \ $\pa \cdot V = 0$, \ as we shall show elsewhere.

\h Let us pass now to the {\em free} case ($A_{\mu} = 0$), when eq.(14a) 
may be written

\

\hfill{$\dopsi \ga_1 \ga_2 + p \psi \ga_0 \; = \; 0 \; ,$
\hfill} (14'a)

\

and admits some simple solutions.$^{12}$ \  Actually, in this case $p$ is
constant [cf. eq.(14c)] and one can choose the {$\ga_{\mu}$} frame so that
 \ $p = m \ga_0$ \ is a constant vector in the direction $\ga_0$. \ Since \ 
$\dox = \psi \ga_0 \psit$, \ it follows that

\

\hfill{$v \: = \: \dis{{1 \over m}} \psi p \psit \:; \;\;\;\; 
\dis{{p \over m}} \: = \: \psi^{-1} v \psit^{-1} \; .$
\hfill} (17)

\

The mean value of $v$ over a zitterbewegung period is then given by the
relation 

\

\hfill{$\lan v \ran_{\rm zbw} \: = \: p/m \: = \:
\psi^{-1} v \psit^{-1}$
\hfill} (18)

\

which resembles the ordinary quantum--mechanical mean value for the 
wave--function $\psi^{-1}$ (recall that in Clifford algebra for any $\psi$ 
it exists its inverse). \ Let us recall, by comparison, that the time
average of $v_{\mu}$ given in eq.(2c) over a zbw period is evidently
equal to $p_{\mu}/m$.   \\ 
\h Let us explicitly stress that, due to the first one [$z \longra \psi \veps$] 
of eqs.(10), the results found by BZ for $z$ are valid as well for $\psi$ in
our formalism.  For instance, for BZ [cf. eq.(1')] it was  \ $v^{\mu} = 
\zb \gabf^{\mu} z \equiv
v^{\mu}_{\rm BZ}$, \ while in the Clifford formalism [cf. eq.(11)] it is \ 
$v \: = \: \psi \ga^0 \psi \: = \: \lan \ga^0 \psit \ga^{\mu} \psi \ran_0 
\ga_{\mu} \: = \: v^{\mu}_{\rm BZ} \ga_{\mu}$. \ As a consequence, $\sig$
refers in general to a cylindrical helix (for the free case) also in our 
formalism.\\
\h Going back to eq.(14'a), by the second one of eqs.(17) we
finally obtain our non-linear (free) Dirac--like equation in the following form:

\

\hfill{$v \cdot \pa \psi \ga_1 \ga_2 \: + \: m \psi^{-1} v \psit^{-1} \psi
\ga_0 \; = \; 0 \; ,$
\hfill} (19)

\

which in the ordinary, tensorial language would write: \ \ 
$i(\bar{\Psi} \gabf^{\mu} \Psi) \pa_{\mu} \Psi \, = \, \gabf^{\mu} p_{\mu}
\Psi$, \ \ 
where $\Psi$ and $\gabf^{\mu}$ are now an ordinary Dirac spinor and
the ordinary Dirac matrices, respectively, and \ ${\hat{p}}_{\mu} \equiv
i \pa_{\mu}$.\\ 

\h In connection with this fundamental equation of motion (19), let us
explicitly notice the following. \ At a classical level, the equation of
motion in the BZ theory was eq.(14a), which held
for the world--line $\sig$.  In other words, eq.(14a) was
valid for {\em one} world--line; on the contrary, eq.(19) is
a {\em field} equation, satisfied by quantities $\psi (x)$ such that 
\ ${\psi (x)}_{\mid \sig} \, = \, \psi (\tau)$. \ A change in interpretation 
is of course necessary when passing from the 
classical to the ``quantum" level: and therefore eq.(19) is now to be 
regarded as valid 
for {\em a congruence} of world lines, that is to say, for a congruence of
stream--lines of the velocity field $V = V(x)$. \ In the quantum case, 
the ``particle"
can follow any of those integral lines, with probability amplitude $\rho$.
 \ \ In this context, it must be recalled that a tentative interpretation of
the Dirac equation within the Clifford algebra approach has been suggested
in ref.$^{20}$, and later in ref.$^{31}$. \ However, in the present paper 
we shall not put forth,
nor discuss, any interpretation of our formalism.\\
\h As we have seen,  eq.(19) will hold for our helical 
motions. Notice moreover that, since [cf. eq.(8)] it is
 \ $\psi = {\r^{1 \over 2}} {\erm^{{\bt \gamma_{5}} /2}} R$, \ in the case 
of the helical
path solution the Lorentz ``rotation" $R$ will be the product of a pure space
rotation and a boost.\\

\h It is important to observe that, if we replace $v$ in eq.(19) by its 
mean value over a zbw period, eq.(18), then we end up with the 
Dirac--Hestenes equation
[{\em i.e.}, the ordinary Dirac equation!], valid now for the center-of-mass
world--line.  In fact, since \ $\lan v \ran_{\rm zbw} \: = \: p/m$, \ eq.(19)
yields

\

\hfill{$p \cdot \pa \psi \ga_1 \ga_2 \: + \: mp \psi \ga_0 \; = \; 0 \; ,$
\hfill} (19')

\

and therefore \ ---if we recall that for the eigenfunctions of $p$ in the DH 
approach$^{20}$ it holds \ $\pa \psi \ga_1
\ga_2 = p \psi$, \ so that $(p \cdot \pa) \psi \ga_1 \ga_2 = p \pa \psi \ga_1
\ga_2$--- \ one obtains:

\

\hfill{$p (\pa \psi \ga_1 \ga_2 + m \psi \ga_0) \; = \; 0 \; ,$
\hfill} (19")

\

which is satisfied once it holds the {\em ordinary} Dirac 
equation (in its Dirac--Hestenes form):

\

\hfill{$\pa \psi \ga_1 \ga_2 \: + \: m \psi \ga_0 \; = \; 0 \; .$
\hfill} (20)

\

\h Let us observe that all the eigenfunctions of $p$ are solutions both
of eq.(19) and of the Dirac equation.\\

\h In conclusion, our non-linear Dirac--like equation (19) is a 
quantum--relativistic equation, that can be regarded as ``sub-microscopic"
in the sense that it refers to the internal motion of a point-like 
``constituent" $\cal Q$. In fact, the density current $\psi \ga^0 \psit$,
relative to the solutions $\psi$ of eq.(19), does oscillates in time in
a helical fashion, in
complete analogy to the initial equation (2c); so that $x$ and $\dot{x}$
refer to $\cal Q$. \ \ On the contrary, the ordinary (linear) Dirac
equation can be regarded as the equation describing the global motion of
the geometrical centre 
of the system ({\em i.e.}, of the whole ``electron"); actually, it has been
obtained from eq.(19) by linearization, that is to say, replacing the 
density current $v \equiv \psi \ga^0 \psit$ by its time
average $p/m$ over the zbw period [cf. eqs.(17)--(18)].\\
\h At last, let us underline that, in the free case,  
eq.(14'a) admits also a trivial solution $\sig_0$, 
corresponding to rectilinear motion. 

\vspace*{1.5 cm}

{\bf 4. \ A VERY SIMPLE SOLUTION}\\  

\h In the free case, the equation

\

\hfill{$\dopsi \ga_1 \ga_2 + p \psi \ga_0 \; = \; 0 $
\hfill} (14'a)

\

admits also a very simple solution (the limit of the ordinary helical
paths when their radius $r$ tends to zero), which ---incidentally--- escaped 
BZ's attention.  \ In fact, let us recall that in this case eq.(14c) implies 
$p$ to be constant, and we were thus able to choose \ 
$p = m \ga_0$. \ As a consequence, quantities $R$ entering any solution $\psi$ 
of eq.(14'a) become pure {\em space} rotations. 
Actually, if solution $\psi$ is essentially a pure space rotation, it holds  \
$\psi \ga_0 \psit = \ga_0 \r$, \ and eq.(14'a) becomes \ $\psi^{-1} \dopsi
= m \ga_1 \ga_2$, \ so that one verifies that

\

\hfill{$\psi \; = \; \r^{1 \over 2} \exp [- \ga_2 \ga_1 m \tau] \; =
 \; \psi (0) \exp [- \ga_2 \ga_1 m \tau]$
\hfill} (21)                      

\

is a (very simple) solution of its. \ \ Moreover, from eq.(14b) it follows

$$v \equiv \dox \; = \; \r \ga_0 $$

and we can set \ $\r = 1$, \ which confirms that our trivial solution (21)
corresponds to rectilinear uniform motion, and that $\tau$ in this case is 
just the proper--time along the particle world--line $\sig_0$. \ (In this case,
of course, the zbw disappears).\\   
\h But, recalling eq.(16), equation (14'a) in the free case may read

$$v \cdot \pa \psi \ga_1 \ga_2 \: + \: m v  \psi \ga_0 \; = \; 0 $$

and finally (using an argument analogous to the one leading to eq.(19"))

$$v (\pa \psi \ga_1 \ga_2 + m \psi \ga_0) \; = \; 0 \; ,$$

which is satisfied once it holds the equation

\

\hfill{$\pa \psi \ga_1 \ga_2 \: + \: m \psi \ga_0 \; = \; 0 $
\hfill} (20')

\

which, as expected, is just the Dirac equation in the Clifford formalism.\\

\h Before going on, we want to explicitly put forth the following observation.
Let us first recall
that in our formalism the (Lorenz force) equation of motion for a charged
particle moving with velocity $w$ in an electromagnetic field $F$ is

\

\hfill{${\dot w} \; = \; {\dis{{e \over m}}} F \cdot w \; .$
\hfill} (22)

\

Now, for {\em all} the free--particle solution of the BZ theory in the Clifford
language, it holds the ``Darboux relation":

\

\hfill{$ {\dot e}_{\mu} \; = \; \Om \cdot e_{\mu} \; , $
\hfill} (23)

\

so that the ``sub--microscopic" point--like object $\cal Q$, moving along 
the helical 
path $\sig$, is endowed [cf. eq.(15)] with the angular--velocity bivector

\

\hfill{$\Om \; = \; {1 \over 2} \, {\dot e}_{\mu} \wedge e^{\mu} \; = \;
{1 \over 2} \, {\dot e}_{\mu} e^{\mu} \; ,$
\hfill} (24)

\

as it follows by recalling that $e_{\mu}$ can always be written, like in 
eq.(7), as \ $e_{\mu} \: = \: R \ga_{\mu} {\til{R}}$. \ \ 
Finally, let us observe that eq.(23) yields in particular \ ${\dot e}_{0} \:
= \: \Om \cdot e_{0}$, \ which is formally identical to eq.(22). \ Thus, the
formal, algebraic way we chose [see eq.(22)] for describing that the system as
a whole possesses a non-vanishing magnetic dipole structure suggests  
that the bivector field $\Om$ may be regarded as a kind of {\em internal}
electromagnetic--like field, which keeps the ``sub--microscopic" object $\cal Q$
moving along the helix.$^{32}$ \  In other words, $\cal Q$ may be considered as 
confining itself along $\sig$ [{\em i.e.}, along a circular orbit, in the 
electron CM], via the generation of the internal, electromagnetic--like
field 

\

\hfill{$F_{\rm int} \; \equiv \; {\dis{{m \over {2 e}}}} \: {\dot e}_{\mu} 
\wedge e^{\mu} \; .$
\hfill} \\

\h We shall further discuss this point elsewhere.

\vspace*{1.5 cm}

{\bf 5. \ ABOUT HESTENES' INTERPRETATION}\\   

\h In connection with our 
new equation (16'), or rather with its (free) form (19), we met solutions
corresponding ---in the free case--- to helical motions with constant
radius $r$; as well as a limiting solution, eq.(21), for $r \ria 0$. \ 
We have seen above 
that the latter is a solution also of the ordinary (free) Dirac equation.\\
\h Actually, the solution of
the Dirac equation for a free electron in its rest frame can be written
in the present formalism as:$^{20}$

\

\hfill{$\psi (x) \; = \; \psi (0) \exp [- \ga_2 \ga_1 m \tau]$
\hfill} (25)

\

which coincides with our eq.(21) along the world--line $\sig_0$.\\ 
\h It is interesting to examine how in refs.$^{20}$, even if confronting
themselves only with the usual (linear) Dirac equation and with eq.(25), 
those authors were led to propose for the electron the existence of 
internal helical motions. \ It was first noticed that, in correspondence with 
solution (25), it is \ ${\dot e}_0
\, = \, 0; \;\; {\dot e}_3 \, = \, 0$, \ so that $e_0, \; e_3$ are constants;
while $e_2, \; e_3$ are {\em rotating}$\: ^{9}$ in the $e_2 e_1$ plane ({\em 
i.e.}, in the spin plane$^{20}$) with the zbw frequency $\om = 2m$ \ [$\hbar =
c = 1$]:

\

\hfill{$e_1 (\tau) \; = \; e_1 (0) \, \cos 2m\tau \: + \: e_2 (0) \, \sin
2m\tau$ \hfill}

\hfill{   \hfill} (26)

\hfill{$e_2 (\tau) \; = \; e_2 (0) \, \cos 2m\tau \: - \: e_1 (0) \, \sin
2m\tau$ \hfill}

\

as follows from eq.(7) with \ $R \: = \: \exp [- \ga_2 \ga_1 m \tau]$. \ \ 
Incidentally, by recalling that in the Clifford formalism the spin bivector 
$S$ is given by

\

\hfill{$ S \; = \; R \ga_2 \ga_1 \til{R} {\dis{\hbar \over 2}} \; = \; e_2 
e_1 {\dis{\hbar \over 2}} \; ,$
\hfill} (27)

\

whilst the angular--velocity bivector $\Om$ is given by eq.(24), one then gets

\

\hfill{$p \cdot v \; = \; \Om \cdot S \; = \; m \; ,$
\hfill} (28)

\

which seems to suggest$^{20}$ the electron rest--mass to have an
(internal)
kinetic origin! Hestenes could do nothing but asking himself (following 
Lorentz$^{33,34}$): what is rotating?\\
\h If something was rotating inside the electron, since $v = e_0$ refers 
{\em in this case} to the electron mean motion, {\em i.e.}, is the 
velocity of the whole electron, in refs.$^{20}$ it was {\em assumed} for the 
velocity of the internal ``constituent" $\cal Q$ the value

\

\hfill{$u \; = \; e_0 - e_2 \; ; \;\;\;\; e_0 \equiv v$
\hfill} (29)

\

which is a light--like quantity.  Eq.(29) represents a null vector since 
${e_0}^2 = +1; \;\; {e_2}^2 = -1$ \ (quite analogously, they could have
chosen  \ $u = e_0 - e_1$). \ Notice that the ordinary Dirac current will
correspond to the average $\bar u$ of velocity $u$ over a zbw period; {\em
i.e.}, due to eqs.(26), to: \ $\bar{u} = e_0 = v$, \ as expected. \ However,
if we set \ $u \equiv \dot{\zeta}$, \ then one gets$^{20}$

\

\hfill{$\zeta (\tau) \; = \; (\erm^{\Om \tau} - 1) R_0 \, + \, \zeta_0 \; ,$
\hfill} (30)

\

which is just the parametric equation of a light--like helix \ $\zeta (\tau) \,
= \, x(\tau) + R(\tau)$ \ centered on the stream--line $\sig_0$ with radius
$R_{\rm H}$ given by [$\om = 2m$]:

\

\hfill{${R_{\rm H}}(\tau) \; = \; \erm^{\Om \tau} R_0 \; = \; 
{\dis{{-e_1 \over \om}}} \; = \; {\dis{{-\dot{u} \over \om^2}}} \; .$
\hfill} (31)

\

The parameters in eqs.(30)-(31) were chosen by Hestenes$^{20}$ in {\em such 
a way} that
the helix diameter equals the Compton wavelength of the electron, and the 
angular momentum of the zbw motion yields the correct electron spin. \  It
is possible, incidentally, that such a motion be also at the origin 
of the electric charge; in any case, we saw that, if the electron is
associated with a clock-wise rotation, then the positron will be associated
with an anti--clock-wise rotation, with respect to the motion direction.\\
\h Choice (29), of course, suits perfectly well with the standard discussions 
about the velocity operator$^{35}$
for the Dirac equation, and as a consequence does naturally allow considering 
that helical motion as the classical analog of zbw. \ Moreover, when making 
recourse to the light-like solutions, the electron spin can be regarded as
totally originating from the zbw motion, since the {\em intrinsic} term 
$\Delta^{\mu \nu}$ entering the BZ theory$^{(10)}$ does {\em vanish} as 
the zbw speed tends to $c$.\\

\h However, such approach by Hestenes, even if inspiring and rich of physical 
intuition, seems rather {\em ad hoc} and in need of a few assumptions 
[particularly with regard to eqs.(29) and (31)]. In our opinion, to get a
sounder theoretical ground, it is to be linked with our eqs.(16'), (19) and the
related discussion above.\\

\h In the original BZ theory, there exist particular initial conditions
yielding {\em as a limiting case} a light--like velocity $v^{\mu} = \zb 
\gabf^{\mu} z$; that is, such that $v^{\mu} v_{\mu} = 0$, as it can be 
checked from eq.(2b). \ [For instance, in ref.$^{24}$ light--like helical
paths have been obtained in correspondence with Majorana (singular) DH spinors
$\psi$]. \ However, in our Clifford formalism, which makes recourse to the
Frenet tetrad, quantity $v$ was bound to be time--like. We may, of course,
allow for a light--like $v$; but, in this case, we have to change the explicit
representation of the velocity vector in the BZ lagrangian: for instance, 
to comply with Hestenes'
assumption (29), we have just to set: \ $v \: = \: \psi \ga_0 \psit - \psi
\ga_2 \psit$, \ so that eq.(12) has to be rewritten accordingly as follows:

\

\hfill{$\Lc \; = \; \lan \psit \dopsi \ga_1 \ga_2 \: + \: p(\dox - 
\psi \ga_0 \psit + \psi \ga_2 \psit) \: + \: eA (\psi \ga_0 \psit 
 - \psi \ga_2 \psit) \ran_0    \; .$
\hfill} (12')

\

\h But choice (29) of ref.$^{20}$ is just one possibility; for example, one
might choose

\

\hfill{$u \; = \; e_0 - e_1 - e_2$
\hfill} (32)

\

and in this case we would get for the rotating point--object $\cal Q$ a
{\em space--like} velocity $u$, whose mean value $\bar u$ would still be

\

\hfill{$\bar{u} \; = \; e_0 \; .$
\hfill}

\

This would correspond in our Clifford formalism to representing the velocity
vector as \ $v \: = \: \psi \ga_0 \psit - \psi \ga_1 \psit - 
\psi \ga_2 \psit$, and modifying the lagrangian (12') accordingly.\\
\h In any case, one may observe that also with the choice (29) the 
light--like
$u$ results from the composition of a time--like velocity $e_0$ with a 
{\em space--like} velocity $e_2$. \ Some interesting work in this direction did 
already appear in refs.$^{36}$, by Campolattaro.

\vspace*{1.5 cm}

{\bf 6. \ FURTHER REMARKS}\\  

\h We mentioned, at the beginning, about 
further methods for introducing ---partially, at least--- 
an helical motion as the classical limit of
the ``spin motion".  We want here to show, at last, how to represent in the
Clifford formalism the {\em extrinsic curvature} approach,$^{16,17}$ 
corresponding to lagrangian (6), due to its interest for the development
of the present work.\\
\h Let us first recall that in classical differential geometry one 
defines$^{23}$ the Frenet frame $\{e_{\mu}\}$ of a non-null curve $\sig$ by
the so--called Frenet equations, which with respect to proper time $\tau$
write [besides \ $\dox = e_0 = v$]:

\

\hfill{$\ddot{x} = {\dot e}_0 = K_1 e^1 \, ; \;\;
{\dot e}_1 = -K_1 e^0 + K_2 e^2 \, ; \;\; {\dot e}_2 = -K_2 e^1 + K_3 e^3 \, 
; \;\; {\dot e}_3 = -K_3 e^2$
\hfill} (33)

\

where the i-th curvatures $K_{\rm i}$ (i=1,2,3) are scalar functions chosen 
in such a way that \ ${e_{\rm j}}^2 = -1$, \ with j=1,2,3. \ Quantity $K_1$
is often called curvature, and $K_2, \; K_3$ torsions (recall that in the 
3--dimensional space one meets only $K_1$ and $K_2$, called curvature and
torsion, respectively). \ Inserting eqs.(33) into eq.(24), we get for the 
Darboux (angular--velocity) bivector:

\

\hfill{$\Om \; = \; K_1 e^1 e^0 \, + \, K_2 e^2 e^1 \, + \, K_3 e^3 e^2 \; ,$
\hfill} (34)

\

so that one can build the following scalar function

\

\hfill{$\Om \cdot \Om \; = \; K_{1}^2 - K_{2}^2 - K_{3}^2 \; = \; 
({\dot e}_{\mu} \wedge e^{\mu}) \cdot ({\dot e}_{\nu} \wedge e^{\nu}) \;\; .$
\hfill} (34')

\

\h At this point, one may notice that the square, $K^2$, of the ``extrinsic
curvature" entering eq.(6) is equal to $-K_{1}^2$, \ so that the lagrangian
adopted in refs.$^{16}$ results ---after the present analysis--- to take
advantage only of the first part,

\

\hfill{${\ddot x}^2 \; = \; {\dot e}_0^2 \; = \; -K_{1}^2 \; ,$
\hfill}

\

of the Lorentz invariant (34').                           
On the contrary, in our formalism the whole invariant $\Om \cdot \Om$
suggests itself as the suitable, complete lagrangian for the problem at issue;
and in future work we shall exploit it, in particular comparing the expected 
results with Plyushchay's.$^{37}$\\  
\h For the moment, let us stress here only the possibly important result 
that the lagrangian \ $\Lc \, = \, \Om \cdot \Om$ does coincide (factors
apart)  along the
particle world--line $\sig$ with the auto--interaction term$^{38}$ 

\

\hfill{${\theta^5} \; (\drm \theta^{\mu} \wedge \theta_{\mu}) \cdot (\drm 
\theta^{\nu} \wedge \theta_{\nu}) $ 
\hfill}

\

of the {\em Einstein--Hilbert lagrangian density} written (in the Clifford 
bundle formalism) in terms of tetrads of 1-form fields $\theta^{\mu}$. \ 
Quantity $\theta^5 \equiv \th^0 \th^1 \th^2 \th^3$ is the volume element.\\
\h Finally, we can examine within our formalism the third approach: that one
utilizing Grassmann variables.$^{19}$  For instance, if we recall that the
Grassmann product is nothing but the external part \ $A_{\rm r} \wedge 
B_{\rm s} \, = \, \lan A_{\rm r} B_{\rm s} \ran_{|{\rm r} - {\rm s}|}$ \
of the Clifford product (where $A_{\rm r}$, $B_{\rm s}$ are a r-vector
and a s-vector, respectively), then the Ikemori lagrangian$^{19}$ can be
immediately translated into the Clifford language and shown to be
equivalent to the BZ lagrangian, apart from the constraint $p^2 = m^2$.\\
\h Further considerations about the solutions of our non-linear, Dirac--like,
new equation and the interesting consequences of the present formalism are
in preparation and will appear elsewhere.\\

\vspace*{1.8 cm}


{\bf 7. ACKNOWLEDGEMENTS}\\

The authors acknowledge the important collaboration of G.D. Maccarrone, 
S. Sambataro, R.M. Santilli and particularly of J. Vaz and G. Salesi. \
They are also grateful to 
Asim O. Barut, A. Campolattaro, R.H.A. Farias, E. Giannetto, R.L. Monaco, 
P. Lounesto, A. Insolia, E. Majorana jr., 
J.E. Maiorino, E.C. de Oliveira, M. Pignanelli, G.M. Prosperi, 
M. Sambataro, Q.G. de Souza  and M.T. Vasconselos for stimulating 
discussions. \ Two of them 
(M.P. and W.A.R.) thank moreover I.N.F.N.--Sezione di Catania, Catania, 
Italy, for kind hospitality during the preparation of the first draft of
this work.


\end{document}